\documentstyle[prl,aps]{revtex}
\input{epsf}
\draft
\begin{document}
\twocolumn[\hsize\textwidth\columnwidth\hsize\csname@twocolumnfalse\endcsname
\title{Localized Structures in Nonlinear Lattices
with Diffusive Coupling and External Driving}

\author{Igor Mitkov, Konstantin Kladko,
and A. R. Bishop}
\address{
Theoretical Division and
Center for Nonlinear Studies,
Los Alamos  National Laboratory,
Los Alamos, NM 87545
}
\date{\today}
\maketitle

\begin{abstract}
We study the stabilization of localized structures by discreteness
in one-dimensional lattices of diffusively coupled nonlinear sites.
We find that in an external driving field
these structures may lose their
stability by either relaxing to a homogeneous state
or nucleating a pair of oppositely moving fronts.
The corresponding bifurcation diagram
demonstrates a cusp singularity.
The obtained analytic results are in good quantitative
agreement with numerical simulations.
\end{abstract}

\narrowtext
\vskip1pc]

The problem of dynamics in discrete nonlinear lattices
arises in diverse physical, biological, and engineering systems.
Among the examples are interaction of charge- and spin-density waves
with impurities in correlated electron
materials~\cite{bishop,floriamazo},
arrays of Josephson junctions~\cite{floriamazo},
calcium release waves in living cells~\cite{calcium},
systems of coupled nonlinear oscillators~\cite{hunt}, {\it etc.}
Of special interest are lattices
with {\it continuous} coupling between the nonlinear sites,
because of their rich dynamical behavior. In Ref.~\cite{mitkov}
such systems have been shown to carry propagating {\it burst waves}.
The role of {\it localized structures} as nucleation embryos in
driven nonlinear systems is of wide importance~\cite{buttiker},
but the influence of lattice discreteness on {\it nucleation and
transport} has received much less attention to date.
We will show in this Letter that lattice discreteness indeed results
in qualitatively new phenomena.

Below we study the stabilization of localized
structures by discreteness in a system consisting of
a lattice of diffusively coupled nonlinear sites.
We investigate the properties of such structures and different
mechanisms of their instabilities. We find a hysteresis effect
in the nucleation of propagating fronts from the localized structures
in an external field.
We obtain our results analytically and confirm them by numerical
simulations of the full system dynamics.
Our system has relaxational dynamics given by
$\partial u/\partial t = - \delta {\mathcal E}/\delta u\,$,
where ${\mathcal E}$ is the system energy functional,
$u$ is the order parameter, and
$\delta/\delta u$ is a variational derivative.
We consider the energy functional of the following form
\begin{equation}
{\mathcal E}[u] \;=\;
\frac{D}{2} \int \left( \frac{\partial u}{\partial x}\right)^2
dx \;+\; \alpha\,\sum_i {\mathcal F}[u_i]\;.
\label{functional}
\end{equation}
Here $D$ is the diffusive coupling, ${\mathcal F}$ is
the discrete nonlinear potential, $\alpha$ is the
potential amplitude, and the sum is over all lattice sites $i\,$.
We study the behavior of system~(\ref{functional})
with bistable dynamics. The details of the potential
shape are not important so long as it has
at least two minima, $u=u_-$ and $u=u_+$
separated by a barrier with a maximum at $u=u_0\,$.
The simplest example of such a potential ${\mathcal F}$
is a fourth degree polynomial ($\phi^4$).
Here we study the sine-Gordon potential, because of its
applicability to modeling the dynamics of charge density
waves~\cite{bishop}. Although this potential
has an infinite number of wells, for the localization problem
studied in this Letter, only two neighboring minima
are relevant. The potential has a form
\begin{equation}
{\mathcal F}[u_i] \;=\; -\cos u_i -E\,u_i\;,
\label{potential}
\end{equation}
where $E$ is the applied external field. The field is required
to make the depths of the potential wells different and thus
allow for the propagation of fronts from less stable to
more stable states.
For potential~(\ref{potential}), these stable states are
$u_- = \arcsin E\,,\,
u_+ = u_- + 2\pi\,,\, u_0 = \pi - u_-\,$.
The system dynamics are then governed by
\begin{equation}
u_t \;=\; \beta u_{xx} \;+\; \sum_i \delta(x-i)(-\sin u + E)\;,
\label{eq2}
\end{equation}
with the space and time rescaled as $x \rightarrow x/d$
($d$ is the distance between sites) and $t \rightarrow t\alpha\,$,
respectively. The renormalized dimensionless
diffusive coupling is $\beta = D/\alpha d^2\,$.

When $\beta \gg 1\,$, the system is in the continuous limit,
where it is described by the overdamped sine-Gordon equation:
$u_t = \beta u_{xx} - \sin u + E\,$.
If one starts with the bulk of the system in the state
$u=u_-\,$, below the barrier maximum $u_0\,$,
and a finite size nucleus of the state $u=u_+\,$, above the maximum,
then, for $E>0\,$, there exists a critical size of the nucleus.
If the nucleus exceeds this critical size,
it breaks into a pair of oppositely moving fronts;
otherwise it relaxes back to the bulk state.
We show that introducing discreteness into the system
can stabilize the critical nucleus as localized structure.

\begin{figure}[h]
\rightline{ \epsfxsize = 10.0cm \epsffile{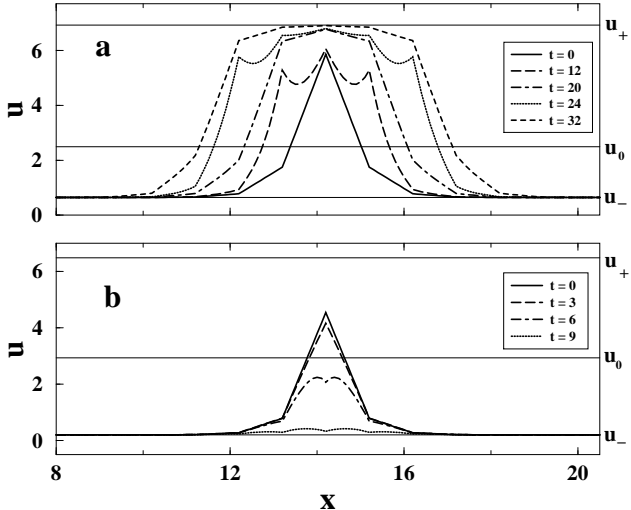}}
\caption{
Successive profiles of the numerical solution $u(x)\,$
of Eq.~(\protect\ref{eq2}), starting from a one-site nucleus.
(a) Nucleation of two oppositely moving burst waves;
$E = 0.6\,,\, \beta = 0.13\,$.
(b) Relaxation to the bulk state $u=u_-\,$;
$E = 0.2\,,\, \beta = 0.17\,$.
For both cases the system length is $L = 30\,$, the number of
grid-points is $300\,$, time step $dt = 2\times 10^{-5}\,$.
\label{fig1}}
\end{figure}

In the discrete regime, $\beta\lesssim 1\,$, the system
demonstrates substantially richer behavior.
Well-separated fronts undergo a {\it pinning-depinning
transition} from stationary kinks to propagating
burst waves, which are periodic in a traveling-wave reference
frame~\cite{mitkov}.
A stabilized nucleus represents a bound pair of kink
and antikink. There exists a hierarchy of nuclei with
differing numbers of sites at the state $u=u_+\,$, above the maximum.
The distance between the kink and the antikink increases
with the number of these sites, and the binding
energy rapidly decays. Since the nucleation is caused by local fluctuations,
the physically most important nuclei are those with small numbers of sites.
As the parameters vary, a nucleus may lose its stability.
Fig.~\ref{fig1} shows two alternative scenaria of one-site nucleus
destabilization:
(a) breaking to a pair of burst waves and
(b) relaxation to the bulk state.

\begin{figure}[h]
\rightline{ \epsfxsize = 10.0cm \epsffile{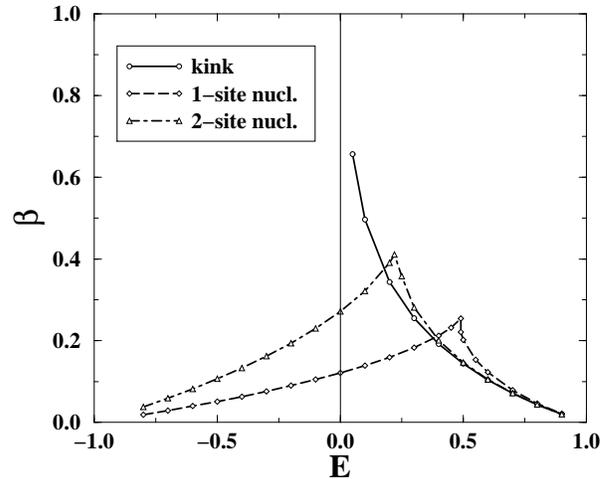}}
\caption{
Numerically obtained bifurcation diagram of kink
and one- and two-site nuclei. Parameters are
as in Fig.~\protect\ref{fig1}.
\label{fig2}}
\end{figure}

The bifurcation diagram of one- and two-site nuclei along with a single
front in the ($E\,, \beta$) plane is shown in Fig.~\ref{fig2}.
The single kink is pinned below its bifurcation line and turns into
a propagating burst wave above the line.
Each nucleus is stable below its bifurcation line, which
consists of two branches.
Crossing the right or left branch upward results in
the nucleation of a pair of burst waves or relaxation to the bulk state,
$u=u_-\,$, respectively.
The two branches intersect at a singular point,
known as a cusp catastrophe~\cite{arnold}.
We see in Fig.~\ref{fig2} that the nucleation of burst waves
from localized nuclei demonstrates {\it hysteresis}, {\it i.e.}
there is a parameter range, where a localized nucleus is stable,
while a single front already propagates.
The bifurcation lines for multi-site nuclei look similar.
The field value at the singular point on each of the lines
goes to $\infty\,$ and $\beta\to 0$, with the number
of sites in the nucleus, due to the increasing separation of
kinks. Note that the hysteresis and the nucleation phenomena
mentioned above happen below the threshold driving strength
($|E| = 1$ in Fig.~\ref{fig2}), at which the effective potential
from Eq.~(\ref{potential}) loses its minima.

For any stationary solution, the dynamic equation~(\ref{eq2})
reduces to the following tridiagonal system of algebraic
equations for $u_i$'s, the values of $u$ at the sites $x_i$'s,
\begin{equation}
\beta\,(u_{i+1} + u_{i-1} - 2 u_i) - \sin u_i + E\;=\; 0\;.
\label{eq3}
\end{equation}
This arises because, in the stationary regime, Eq.~(\ref{eq2})
between the sites
turns into a one-dimensional Laplace equation
with linear solutions.

We notice that the maximum value of $\beta$ on the one-site
nucleus bifurcation line in Figure~\ref{fig2}, is $\sim 0.25\,$,
{\it i.e.} deep
in the discrete (small $\beta$) regime, which ends near
$\beta \sim 1\,$. It is, therefore, natural
to study the stability of the nucleus, taking $\beta$ as a
small parameter of the analysis.

\vspace{-0.5cm}

The following three sites are the key elements of our analysis:
the site of the nucleus, $i=n\,$, and its two
nearest neighbors, $i=n \pm 1\,$, with the corresponding field values
$u_n$ and $u_{n\pm 1}\,$. The symmetry of the nucleus implies
that $u_{n+1}=u_{n-1}\equiv u^*\,$.
The neighboring sites are both equivalent to the
``front site'' on a single kink. They are most active in a sense
that they are the first candidates to cross the threshold
value $u_0$, if the stationary solution becomes unstable.
It can be shown (cf. results of Ref.~\cite{mitkov} for
a single kink) that the field value at the other sites
decays with the distance from the front site down to $u_-\,$,
as $u_{n\pm k}-u_-\sim\beta^{k-1}\,$ (for $k=2,3,\ldots$).
Consider the stability analysis to first order in $\beta\,$.
Then only the abovementioned three sites make a nontrivial contribution
to the system stability properties, which leaves only two
independent variables, $u_n$ and $u^*\,$. Using the corresponding
two equations from the system~(\ref{eq3}), we reduce them to one
equation for the ``active'' value $u^*$
\begin{eqnarray}
g(u^*) \;&\equiv&\; 2\beta\,(u_- - u^*) - 2\sin u^* + 3E
\nonumber\\
&-&\sin\left[2 u^* - u_- + \frac{1}{\beta}(\sin u^* - E)\right]
\;=\; 0\;.
\label{first_order}
\end{eqnarray}

\begin{figure}[h]
\rightline{ \epsfxsize = 9.0cm \epsffile{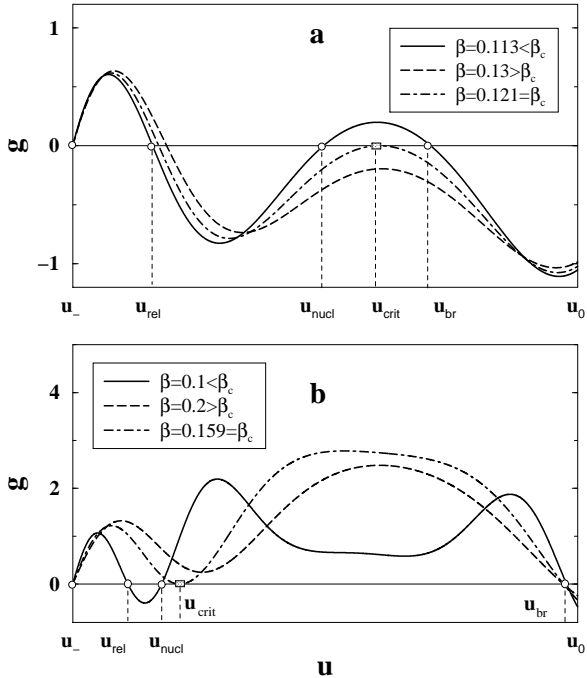}}
\caption{
Solution of Eq.~(\protect\ref{first_order}).
(a) $E=0.6\,,\, \beta_c = 0.121\,$; (b) $E=0.2\,,\, \beta_c = 0.159\,$.
\label{fig3}}
\end{figure}

The solution of Eq.~(\ref{first_order}) is represented graphically
in Fig.~\ref{fig3}. The case of the right (breaking) branch
of the one-site nucleus bifurcation line (see Fig.~\ref{fig2}) corresponds
to Fig.~\ref{fig3}(a). Then below
the bifurcation line, the function $g(u)$ in Eq.~(\ref{first_order})
is given by the solid line, and the equation has four solutions
(open circles in the figure).
To determine the stability of these solutions, one has to
linearize system~(\ref{eq3}) and find the eigenvalues of
the obtained linear problem.
We note that the maximum eigenvalue $\lambda_{max}$
is responsible for the stability of the most active front site
with $u= u^*\,$. This means that near the bifurcation line
only $\lambda_{max}$ can change sign, whereas all other
eigenvalues remain negative. To find $\lambda_{max}$, we
linearize Eq.~(\ref{first_order}) and conclude that the 1st
and the 3rd solutions for $u^*$ have $\lambda_{max} < 0\,$,
while the 2nd and the 4th solutions have $\lambda_{max} > 0\,$.
The two former solutions are, therefore, the stable nodes,
and the two latter solutions are saddles with one unstable
direction in the phase space.

The 1st solution, however, is trivial in the sense that all
sites are in the bulk state $u=u_-\,$. This leaves
the single nontrivial stable solution, $u_{nucl}$
in Fig.~\ref{fig3}. As $\beta$ increases, the stable solution
and the right saddle, $u=u_{br}\,$, approach each other and
eventually merge at $\beta=\beta_c\,$.
This occurs when the maximum of curve $g(u)$ in Fig.~\ref{fig3}(a)
touches the
line $g=0\,$, at $u=u_{crit}$. Apparently the two corresponding maximum
eigenvalues both become zero, which implies a saddle-node bifurcation.
Further
increase of $\beta$ leaves the system without a stationary solution,
and the nucleus breaks, leading to the formation of a pair of
oppositely moving burst waves [Fig.~\ref{fig1}(a)].

Stability analysis of the left (relaxation) branch of the
bifurcation line (see Fig.~\ref{fig2}) can be made analogously,
using Fig.~\ref{fig3}(b). The difference is that now
the stable solution merges with the left (relaxation) saddle, $u=u_{rel}\,$,
and the nucleus relaxes to the bulk state [Fig.~\ref{fig1}(b)].

\begin{figure}[h]
\rightline{ \epsfxsize = 10.0cm \epsffile{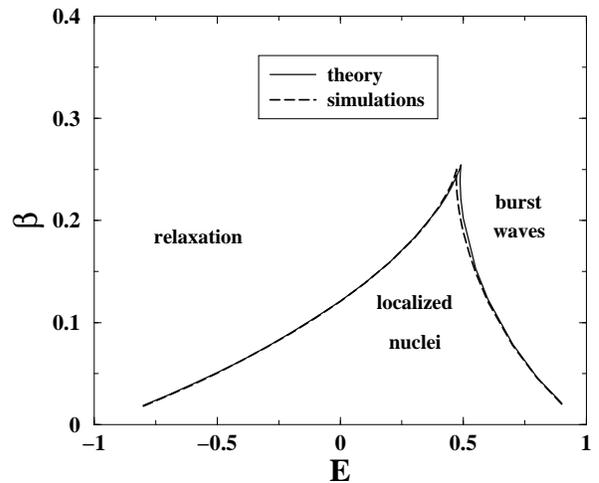}}
\caption{
Comparison between analytical and numerical bifurcation
lines for a one-site nucleus. Parameters as in Fig.~\protect\ref{fig1}.
\label{fig4}}
\end{figure}

It appears that the first-order analysis is insufficient
for the quantitative comparison of our theory with numerical
simulations of Eq.~(\ref{eq2}). To improve our predictions,
we have developed the stability analysis to
second order in $\beta\,$. In this case one has to consider
the nucleus and two of its neighbors from each side. Symmetry
of the problem reduces the number of independent variables
from five to three, and we deal with three corresponding equations from
system~(\ref{eq3}). We then perform the same procedure
as for the first-order stability analysis
described above.
In Fig.~\ref{fig4} we plot the resulting bifurcation line for the single-site
nucleus along with the results of the simulations from Fig.~\ref{fig2}.
We see in Fig.~\ref{fig4} that our theoretical prediction
is in good quantitative agreement with the simulations.

\vspace{-0.5cm}
\vspace{-0.5cm}

We now describe the physical mechanisms of the obtained
phenomena of relaxation and break-up of localized
nuclei. In the parameter regime below the bifurcation line,
the energy functional ${\mathcal E}$, Eq.~(\ref{functional}),
possesses a minimum in the
functional space, which corresponds to a stable nucleus.
Separatrices connect this minimum to the two neighboring 
unstable solutions (saddles). If one starts with an initial
condition in either of these saddles slightly
perturbed in the direction of the minimum, the system
evolves along the separatrix until it relaxes
at the minimum. If one parameterizes the position on
the separatrix with an arclength $s$, defined as
$ds^2 = dt^2 \int (\partial u/\partial t)^2 dx\,$,
the system dynamics take a simple gradient form,
$ds/dt = - d{\mathcal E}(s)/ds\,$.

\begin{figure}[h]
\rightline{ \epsfxsize = 10.0cm \epsffile{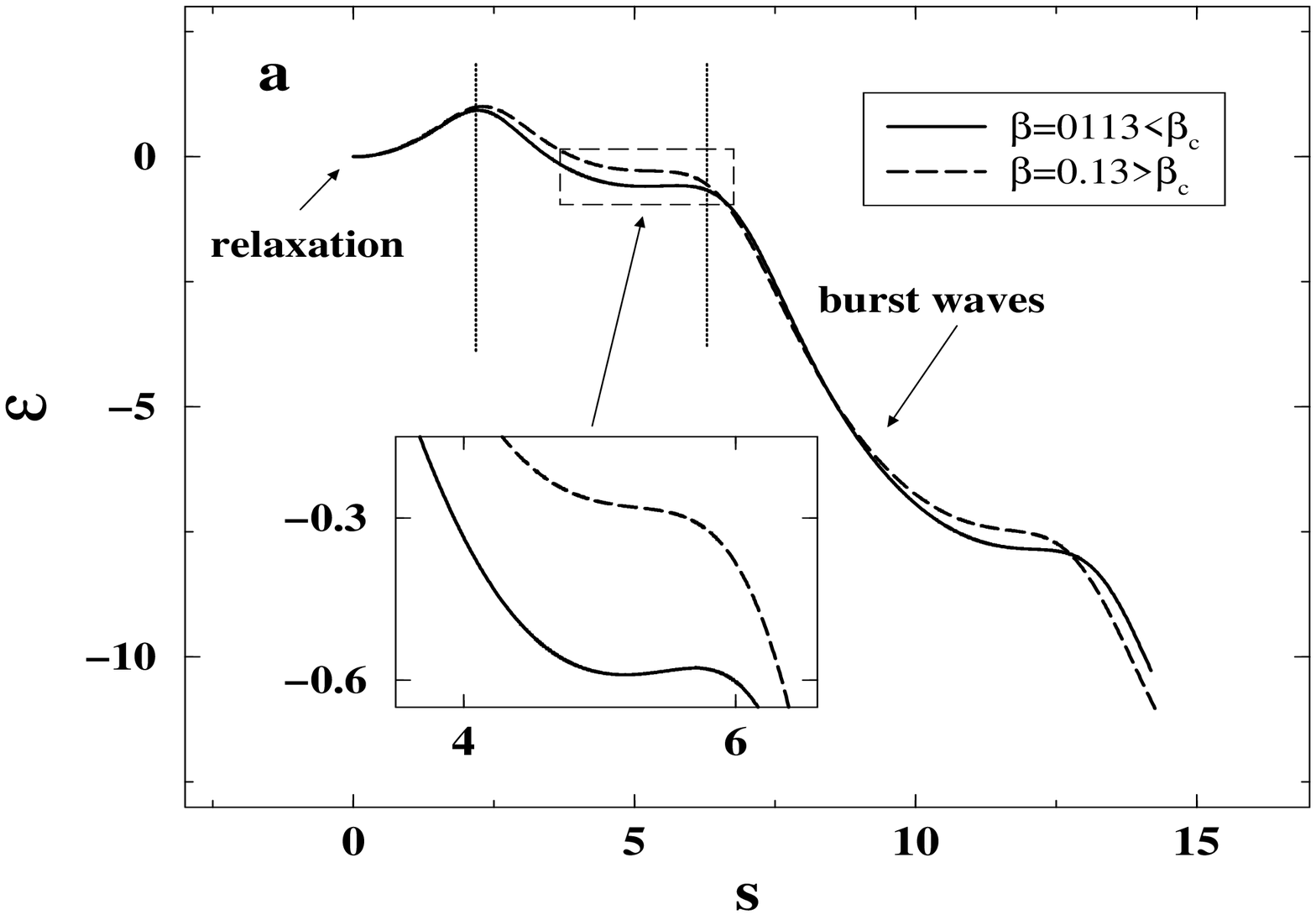}}
\rightline{ \epsfxsize = 10.0cm \epsffile{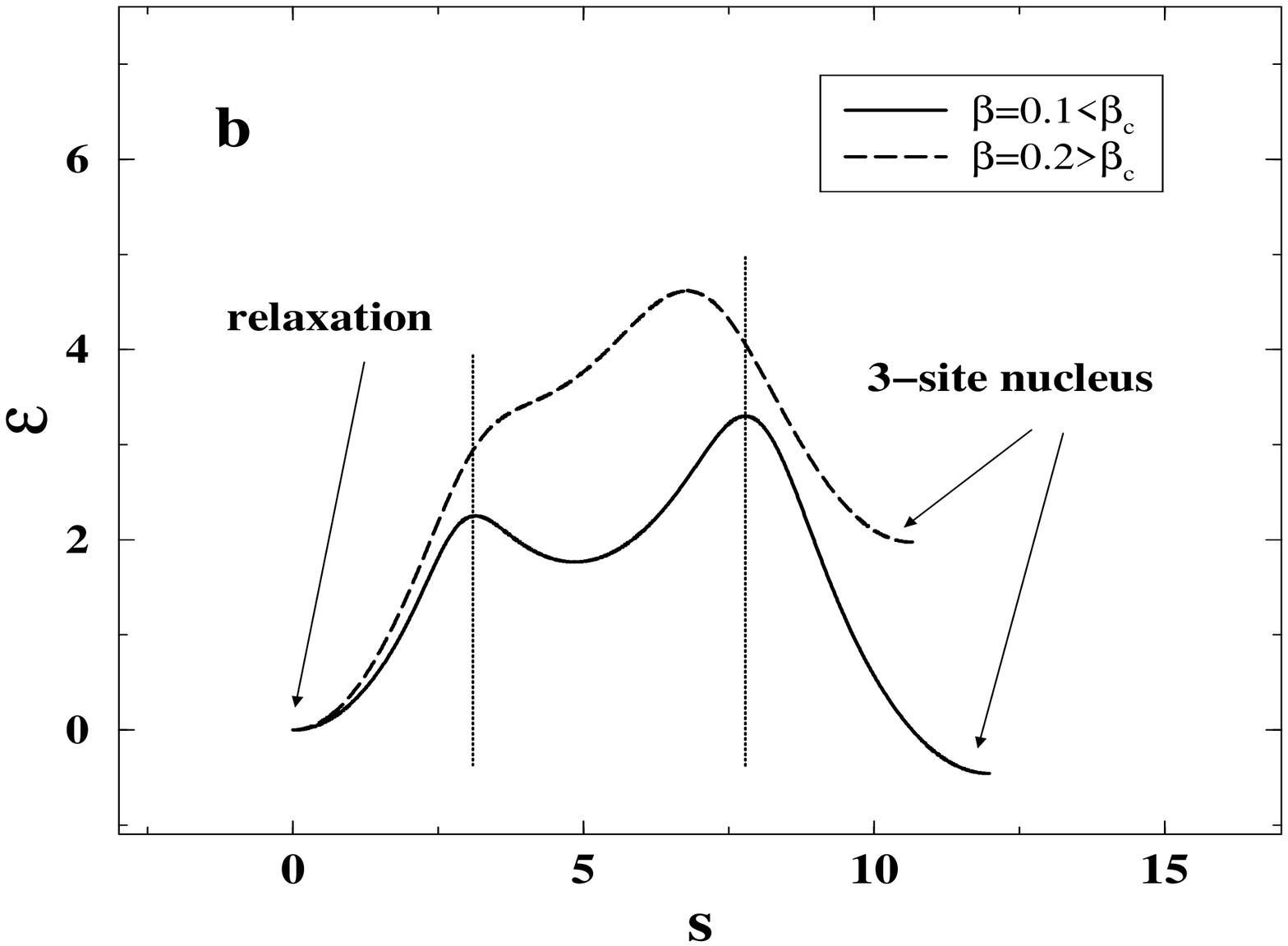}}
\vspace{0.5cm}
\caption{
Energy ${\mathcal E}$ vs arclength $s$ near
(a) break-up branch ($E=0.6$) and (b) relaxation
branch ($E=0.2$) of the bifurcation line of a one-site nucleus.
\label{fig5}}
\end{figure}

In Fig.~\ref{fig5} we plot the dependence of energy
${\mathcal E}$ on $s\,$ from numerical simulations.
Fig.~\ref{fig5}(a) shows ${\mathcal E}(s)$ near the break-up
branch of the bifurcation line. We see in the figure that, for
the stable nucleus (solid line), the energy indeed has
a minimum and two neighboring maxima, corresponding to the
saddles of the dynamics. When the nucleus loses stability (dashed line),
the minimum merges with the right (break-up) saddle. Then the nucleus
turns into a pair of burst waves and the energy begins
monotonically decreasing. If we start with an initial condition
slightly to the right
of the break-up saddle then the system develops a pair of burst waves,
even in the stable nucleus regime. This is the manifestation
of the nucleation hysteresis effect, since the bound nucleus is still
stable in the parameter range for which a well-separated front
already bursts and propagates. If, on the other hand, we start
to the left of the relaxation saddle, then the system will relax
to the homogeneous bulk state $u=u_-$ for both stable and unstable
nuclei.

\vspace{-0.5cm}

The system behavior near the relaxation branch of the bifurcation line
is different, see Fig.~\ref{fig5}(b).
As $\beta$ crosses the critical value, the minimum
of the functional merges with the left (relaxation) saddle,
and the system relaxes to the homogeneous bulk state. The same thing
happens if we start to the left of this saddle, for the regime
of parameters where the nucleus is stable. However, if we start
slightly to the right of the break-up saddle, then the nucleus
at first starts breaking, but then relaxes to the three-site
nucleus, rather than developing into the pair of separated burst waves.

In conclusion, we have demonstrated that nonlinear lattices
with diffusive coupling possess localized structures, in
certain parameter regimes. We have shown that these
structures can be destroyed in two alternative scenaria:
({\it i}) break-up into a pair of oppositely propagating
burst waves, and ({\it ii}) relaxation to a homogeneous
state. We have found hysteresis in the
burst-wave nucleation from a localized embryo, appearing as
the difference between the stability thresholds of
nucleus break-up and burst-wave propagation.
We have given a theoretical description of these phenomena
in terms of the energy functional of our system.
The predictions of our theory are in a good quantitative
agreement with numerical simulations of the full
system. Our interesting directions for future research:
the influence of the localized structures
on thermal nucleation~\cite{buttiker}, dynamics of localized structures
in corresponding two- and three-dimensional systems, experimental studies
of stable and unstable nuclei in materials, {\it etc}.
We note that, though we have studied
lattices with continuous coupling, the stability analysis
has been performed for stationary localized structures from Eq.~(\ref{eq3}).
Therefore, the obtained stability properties of these
structures should remain intact for
completely discrete systems, such as arrays of
Josephson junctions~\cite{floriamazo} or lattices of nonlinear
oscillators~\cite{hunt}.

We appreciate fruitful discussions with D. K. Campbell and J. E. Pearson.
This research is supported by the Department of Energy under contract
W-7405-ENG-36.

\end{document}